\documentclass{elsart}
\begin{document}
\input epsf

\begin{frontmatter}

\title{Bose-Einstein Correlations of Charged Kaons\\ 
in Central Pb+Pb Collisions at $E_{beam} = 158 \, AGeV$} 

\collab{The NA49 Collaboration}

\noindent
S.V.~Afanasiev$^{9}$,T.~Anticic$^{21}$, B.~Baatar$^{9}$,D.~Barna$^{5}$,
J.~Bartke$^{7}$, R.A.~Barton$^{3}$, M.~Behler$^{15}$,
L.~Betev$^{10}$, H.~Bia{\l}\-kowska$^{19}$, A.~Billmeier$^{10}$,
C.~Blume$^{8}$, C.O.~Blyth$^{3}$, B.~Boimska$^{19}$, M.~Botje$^{1}$,
J.~Bracinik$^{4}$, R.~Bramm$^{10}$, R.~Brun$^{11}$,
P.~Bun\v{c}i\'{c}$^{10,11}$, V.~Cerny$^{4}$, O.~Chvala$^{17}$,
J.G.~Cramer$^{18}$, P.~Csat\'{o}$^{5}$, P.~Dinkelaker$^{10}$,
V.~Eckardt$^{16}$, P.~Filip$^{16}$,
Z.~Fodor$^{5}$, P.~Foka$^{8}$, P.~Freund$^{16}$,
V.~Friese$^{8,15}$, J.~G\'{a}l$^{5}$,
M.~Ga\'zdzicki$^{10}$, G.~Georgopoulos$^{2}$, E.~G{\l}adysz$^{7}$, 
S.~Hegyi$^{5}$, C.~H\"{o}hne$^{15}$, G.~Igo$^{14}$,
P.G.~Jones$^{3}$, K.~Kadija$^{11,21}$, A.~Karev$^{16}$,
V.I.~Kolesnikov$^{9}$, T.~Kollegger$^{10}$, M.~Kowalski$^{7}$, 
I.~Kraus$^{8}$, M.~Kreps$^{4}$, M.~van~Leeuwen$^{1}$, 
R.~Lednicky$^{16}$,
P.~L\'{e}vai$^{5}$, A.I.~Malakhov$^{9}$, S.~Margetis$^{13}$,
C.~Markert$^{8}$, B.W.~Mayes$^{12}$, G.L.~Melkumov$^{9}$,
C.~Meurer$^{10}$,
A.~Mischke$^{8}$, M.~Mitrovski$^{10}$, 
J.~Moln\'{a}r$^{5}$, J.M.~Nelson$^{3}$,
G.~P\'{a}lla$^{5}$, A.D.~Panagiotou$^{2}$,
K.~Perl$^{20}$, A.~Petridis$^{2}$, M.~Pikna$^{4}$, L.~Pinsky$^{12}$,
F.~P\"{u}hlhofer$^{15}$,
J.G.~Reid$^{18}$, R.~Renfordt$^{10}$, W.~Retyk$^{20}$,
C.~Roland$^{6}$, G.~Roland$^{6}$, A.~Rybicki$^{7}$, T.~Sammer$^{16}$,
A.~Sandoval$^{8}$, H.~Sann$^{8}$, N.~Schmitz$^{16}$, P.~Seyboth$^{16}$,
F.~Sikl\'{e}r$^{5}$, B.~Sitar$^{4}$, E.~Skrzypczak$^{20}$,
G.T.A.~Squier$^{3}$, R.~Stock$^{10}$, H.~Str\"{o}bele$^{10}$, 
T.~Susa$^{21}$, I.~Szentp\'{e}tery$^{5}$, J.~Sziklai$^{5}$,
T.A.~Trainor$^{18}$, D.~Varga$^{5}$, M.~Vassiliou$^{2}$,
G.I.~Veres$^{5}$, G.~Vesztergombi$^{5}$,
D.~Vrani\'{c}$^{8}$, A.~Wetzler$^{10}$, C.~Whitten$^{14}$,
I.K.~Yoo$^{8,15}$, J.~Zaranek$^{10}$, J.~Zim\'{a}nyi$^{5}$

\vspace{0.5cm}
\noindent
$^{1}$NIKHEF, Amsterdam, Netherlands. \\
$^{2}$Department of Physics, University of Athens, Athens, Greece.\\
$^{3}$Birmingham University, Birmingham, England.\\
$^{4}$Comenius University, Bratislava, Slovakia.\\
$^{5}$KFKI Research Institute for Particle and Nuclear Physics, 
Budapest, Hungary.\\
$^{6}$MIT, Cambridge, USA.\\
$^{7}$Institute of Nuclear Physics, Cracow, Poland.\\
$^{8}$Gesellschaft f\"{u}r Schwerionenforschung (GSI), 
Darmstadt, Germany.\\
$^{9}$Joint Institute for Nuclear Research, Dubna, Russia.\\
$^{10}$Fachbereich Physik der Universit\"{a}t, Frankfurt, Germany.\\
$^{11}$CERN, Geneva, Switzerland.\\
$^{12}$University of Houston, Houston, TX, USA.\\
$^{13}$Kent State University, Kent, OH, USA.\\
$^{14}$University of California at Los Angeles, Los Angeles, USA.\\
$^{15}$Fachbereich Physik der Universit\"{a}t, Marburg, Germany.\\
$^{16}$Max-Planck-Institut f\"{u}r Physik, Munich, Germany.\\
$^{17}$Institute of Particle and Nuclear Physics, Charles University, 
Prague, Czech Republic.\\
$^{18}$Nuclear Physics Laboratory, University of Washington, 
Seattle, WA, USA.\\
$^{19}$Institute for Nuclear Studies, Warsaw, Poland.\\
$^{20}$Institute for Experimental Physics, University of Warsaw, 
Warsaw, Poland.\\
$^{21}$Rudjer Boskovic Institute, Zagreb, Croatia.\\

\begin{abstract} 
  Bose-Einstein correlations of charged kaons were measured near 
mid-rapidity in central Pb+Pb collisions at 158 A$\cdot$GeV by the 
NA49 experiment at the CERN SPS. Source radii were extracted using 
the Yano-Koonin-Podgoretsky and Bertsch-Pratt parameterizations. 
The results are compared to published pion data.  The measured 
$m_\perp$ dependence for kaons and pions is consistent with collective 
transverse expansion of the source and a freeze-out time of about 9.5 $fm$.
\end{abstract} 

\end{frontmatter}

\section{Introduction} 

  Momentum correlations between pions have been widely used to study 
the space-time extent of the emitting source in nucleus-nucleus collisions. 
For identical bosons the symmetry requirement of Bose--Einstein (BE) 
statistics results in an enhanced production of bosons with small 
momentum difference. This effect was first observed by Goldhaber, 
 Goldhaber, Lee and Pais (GGLP) \cite{gol60} for pairs of like-charge 
pions at small opening angles. Kopylov and Podgoretsky \cite{kop72} 
noticed the deep analogy with Hanbury-Brown and Twiss (HBT) space--
time correlations \cite{han56} of the classical electromagnetic fields used
 in astronomy for interferometric measurement of star angular radii and 
developed the basic methods of momentum interferometry in particle and 
nuclear collisions. This technique represents an important tool for 
investigating the underlying reaction mechanism as far as it influences 
the decoupling configuration. For example,  an unusually long decoupling 
time was suggested as one of the signatures of quark-gluon plasma 
(QGP) formation \cite{pra86,ber88}.

 In this paper, we present data from an experimental study of BE 
correlations between charged kaons emitted in central collisions of Pb 
nuclei at a beam energy of 158 A$\cdot$GeV with a fixed Pb target. 
We compare the results with those from earlier 
analyses of pion data for the same 
reaction. It was argued \cite{gyu90} that correlations of kaons and pions 
might differ due to different contributions from resonance decays during 
their propagation toward freeze-out. A comparison could help to 
distinguish between a QGP deflagration and a conventional production 
mechanism based on independent nucleon-nucleon interactions.  
However, rescattering between the final hadrons may smear out the 
differences and obscure what happened in the earlier stages of the 
collision.

\begin{figure}[t!] 
 \begin{center} 
  \leavevmode 
  \epsfxsize=14 cm 
  \epsfbox{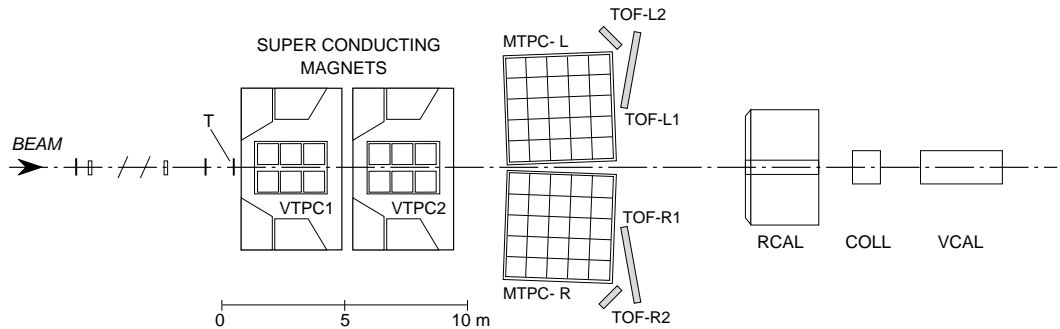} 
 \end{center} 
 \caption{A schematic overview of the NA49 experimental apparatus 
showing the target T, the VTPCs within the superconducting magnets, 
the MTPCs and the TOF walls. Calorimeter VCAL provides the 
centrality trigger.} 
 \label{figure1} 
 \vspace{0.1cm} 
\end{figure} 

 A probably more important aspect of kaons is associated with their 
higher mass. The particle source resulting from a collision between 
heavy nuclei is rapidly expanding in longitudinal and transverse 
directions. This manifests itself in a dependence of the extracted 
correlation parameters on the transverse mass $m_\perp$ of the 
particles, which form the pair \cite{hei96b}. Therefore, kaons and 
pions probe the situation in different kinematic regimes. Assuming 
that the decoupling surface coincides for both particle species, the 
kaons extend the $m_\perp$ range of the pion data and improve the 
sensitivity of comparisons with reaction models.

 Because of their much smaller abundance a study of kaon correlations 
requires higher event numbers and a significantly improved particle 
identification compared to the pion case. The only published results on 
kaon correlation in Pb+Pb collisions were obtained by experiment NA44 
at the CERN SPS \cite{bea01}. The NA49 spectrometer \cite{NIM} includes 
a high-resolution time-of-flight (TOF) detector with acceptance at 
mid-rapidity which allows to select a relatively clean kaon sample. 
Data obtained from an analysis of approximately 1.7 million central as 
well as from 2 million semi-central events will be shown.  The 
K$^+$K$^+$ and K$^-$K$^-$ correlation functions are obtained in 
terms of the Bertsch-Pratt (BP) \cite{pra86,ber88,cha95,pod83} and 
Yano-Koonin-Podgoretsky (YKP) \cite{cha95,pod83,yan78} coordinate frames. 
They are corrected for Coulomb effects. The extracted parameters - 
correlation radii, freeze-out duration -  will be compared with those 
obtained for pions \cite{bea01,app98,ros02,ada02,ant01} in the same central 
rapidity range.

\section{Experiment and Data} 

The 158 A$\cdot$GeV Pb ions from the CERN SPS were measured in 
position and direction by beam position detectors before hitting a Pb foil 
target of 224(336) mg/cm$^2$ thickness (1(2.5) \% interaction probability)
during the 1995/1996(2000) data taking periods.
 
The trigger condition was derived from the signal detected in the veto 
calorimeter (VCAL, see Fig.~\ref{figure1}) which intercepts mainly 
projectile spectators 25 m downstream of the target. For the central data 
sample the 5\% most central interactions were selected. In a geometrical 
picture this corresponds to impact parameters smaller than 3.5 fm. Data 
taken during different run periods contribute 380 000 (1995), 400 000 
events (1996) and 911 000 events (2000). From the 
year 2000 run an additional sample corresponding to semi-central events 
(centrality 5\% - 20\%) was available and analysed.

\begin{figure}[t!] 
 \begin{center} 
  \leavevmode 
  \epsfxsize=14 cm 
  \epsfbox{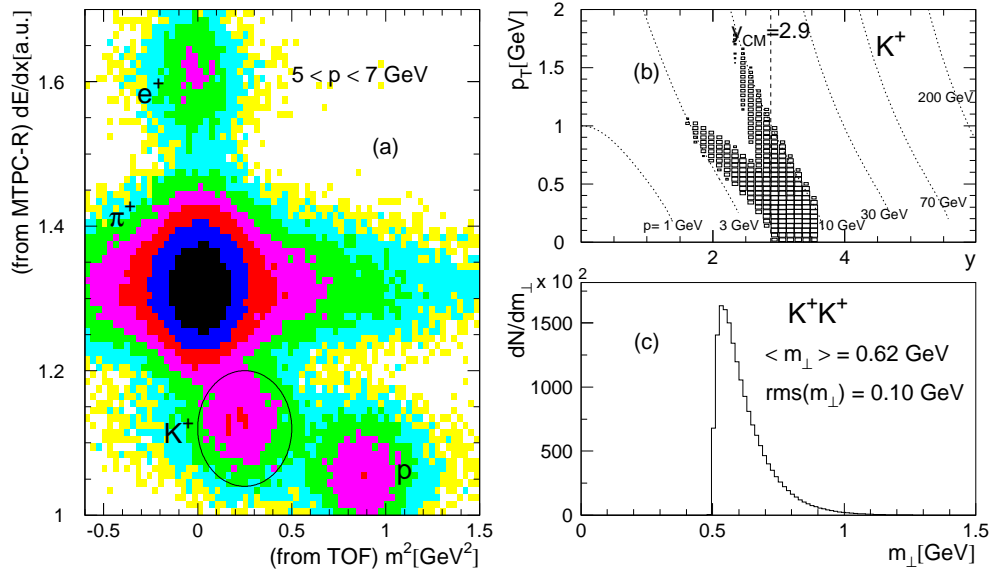} 
 \end{center} 
 \caption{(a) Particle identification by combining TPC and TOF
information. The circled region indicates the selected kaons. 
(b) TOF acceptance region in rapidity $y$ and transverse momentum 
$p_T$ for single K$^+$. Lines of constant momenta are shown by the 
dotted curves and the dashed line indicates the center-of-momentum 
rapidity $\mathrm{y_{CM}}$. (c) $m_\perp$-distribution for 
K$^+$K$^+$ pairs.} 
 \label{figure2} 
 \vspace{0.5cm} 
\end{figure} 

 The NA49 spectrometer (described in detail in \cite{NIM}) consists of 
two superconducting magnets (total bending power of 7.5 Tm) with TPCs 
(time projection chambers) inside the magnetic field, two additional TPCs 
downstream of the magnets, and several TOF (time-of-flight) walls on 
both sides of the beam (Fig.~\ref{figure1}).  
The momentum vectors of charged particles are determined from their 
trajectories as measured in the TPCs and the position of the interaction 
point. The accuracy of the four-momentum difference\footnote{Note that 
the units are chosen such that $c=\hbar=1$} $q$ of a particle pair is of 
the order of 3.5 MeV in the momentum range of interest.

For particle identification a combination of the specific energy loss 
$dE/dx$ measured in the TPCs MTPC-R/L with the mass information 
obtained from the TOF walls TOF-R1/L1 was used. The method is 
illustrated by Fig.~\ref{figure2}a for a typical momentum range. From 
Fig.~\ref{figure2}a it is clear that no satisfactory kaon identification 
can be derived from either the specific ionization or the TOF signal 
alone. Therefore we restrict the phase space used in the present 
study to the acceptance of the TOF walls. Particles inside the region
around the kaon peak (see Fig.~\ref{figure2}a) are selected as 
kaons. The corresponding population 
in the rapidity ($y$) and transverse momentum ($p_T$) plane is 
displayed in Fig.~\ref{figure2}b for the K$^+$ 
mesons; for K$^-$ it is nearly identical. From fits of sums of Gaussian
distributions for the different particle species the purity of the kaon
sample is estimated to be about 93\%, resulting in a purity of 86\% for 
pairs. The distribution of the 
average transverse mass\footnote{$m_\perp=\sqrt{m_K^2+K_T^2}$, where 
$K_T$ is the average transverse momentum of the kaons in the pair.} 
$m_\perp$ of accepted K$^+$K$^+$ pairs is shown in Fig.~\ref{figure2}c. 
Due to statistical limitations a subdivision into $K_T$ intervals was not
possible. Because of slightly different detector efficiencies 
the weighted means of the correlation functions obtained from the 
1995, the 1996 data and the 2000 data were used for the BE 
correlation analysis. Details of the analysis can be found 
in \cite{yoo01}.

\section{Bose-Einstein Correlation Analysis} \label{sec:hbt} 

The correlation function $C_2( {p}_1, {p}_2)$ is defined as the 
ratio of the probability of observing particles with four-momenta 
$ {p}_1$ and $ {p}_2$ simultaneously in one event divided by the 
probability for pairs of independent particles with the same single- particle 
phase space distribution. Experimentally, the raw correlation 
function is obtained as 
\begin{eqnarray} \label{c2_define} 
C_2^{raw}({ {q}})={\mathcal{N}}\frac{S({ {q}})}{B({ {q}})} 
\end{eqnarray} 
where $S( {q})$ denotes the yield of particle pairs from the same 
event with the four-momentum difference ${q}= {p}_1- {p}_2$. 
$B( {q})$ is the uncorrelated reference generated by the event-mixing 
method and the factor $\mathcal{N}$ is used to normalize the correlation 
function to be unity at large relative momenta (200 - 2000 MeV).
 
\begin{figure} 
 \begin{center} 
  \leavevmode 
  \epsfxsize=14cm 
  \epsfbox{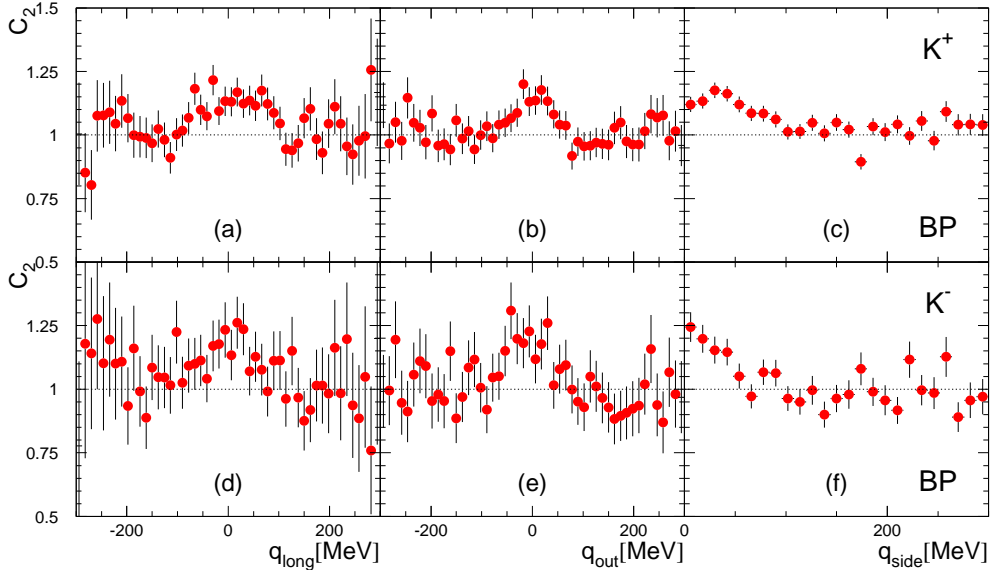} 
 \end{center} 
 \caption{1-dimensional projections on $q_i$ of the 3-dimensional raw 
correlation functions Eq.~\ref{c2_define} with $q_j\le 36$ MeV ($j\ne i$) 
in the BP parameterization. Top and bottom rows are for $K^+K^+$ 
and for $K^-K^-$, respectively.} 
 \label{figure3} 
 \vspace{0.5cm} 
\end{figure} 

Assuming an expanding Gaussian source in a representation given by 
Yano-Koonin-Podgoretsky \cite{yan78} and Heinz \cite{hei96} that is 
appropriate to explicitly take into account collective expansion, 
the correlation due to quantum statistics can be parameterized in terms 
of the components of $q$ : 
\begin{eqnarray} \label{ykp} 
C_2&(&q_\perp,q_\parallel,q_0)=1+\lambda exp[-q_\perp^2R_\perp^2
\nonumber\\ 
&-&\gamma_{YK}^2(q_\parallel-\beta_{YK}q_0)^2R_\parallel^2 
-\gamma_{YK}^2(q_0-\beta_{YK}q_\parallel)^2R_0^2], 
\end{eqnarray} 
where $q_\perp=\sqrt{(\Delta p_x)^2+(\Delta p_y)^2}$ is the transverse 
momentum difference, $q_\parallel=\Delta p_z$ is the longitudinal 
momentum difference, $q_0=E_1-E_2$ is the energy difference, 
$\lambda$ is the correlation intensity parameter and $\beta_{YK}$
 is the ``Yano-Koonin velocity'' which describes the source's 
longitudinal collective motion in the given rapidity range. The radii $R_i$, 
where $i = \perp, \parallel, 0$ denote the transverse, longitudinal and 
temporal extent of the coherence lengths, respectively.

Although the YKP parameters have the most straightforward physical 
interpretation \cite{hei96a}, this parameterization can be ill-defined in 
some kinematic regions \cite{tom99}. Therefore, we use the 
Bertsch-Pratt parameterization in addition. For cylindrically symmetric 
systems and the midrapidity range it reads with vanishing cross terms \cite{cha95} : 
\begin{eqnarray} \label{bp} 
C_2&(&q_{side},q_{long},q_{out})=\nonumber\\ 
&1&+\lambda exp[-q_{side}^2R_{side}^2-q_{long}^2R_{long}^2
-q_{out}^2R_{out}^2], 
\end{eqnarray} 
where $q_{long}$ is identical with $q_\parallel$ defined above, and 
($q_{out}$, $q_{side}$) are the components of the transverse 
3-momentum difference in outward direction and perpendicular to it. 
As reference frame, a pairwise defined co-moving system would be 
appropriate. In view of the limited rapidity range considered here it is 
approximated by the so-called fixed local center-of-mass system, 
thus having the same frame for all particle pairs. Projections of
the raw correlation function of $K^+K^+$ and $K^-K^-$ pairs are
shown in Fig.~\ref{figure3} in BP-variables.

In order to extract the quantum-statistical effects, the counter-acting 
influence of the Coulomb repulsion as well as the dilution by 
misidentified pairs must be taken into account, while the effect 
of the strong interaction can be neglected \cite{sin98}. This leads to 
the ansatz:
\begin{eqnarray} \label{fitfunc}
C_2^{raw} = ( 1 - p ) + p \cdot C_2(q) \cdot A(q) ,
\end{eqnarray}
where $p = 0.86$ denotes the purity of the pairs, $A(q)$ represents the
Coulomb repulsion factor and $C_2(q)$ is the correlation from quantum 
statistics parameterized by  Eqs.~\ref{ykp} or \ref{bp}. Because of the 
considerable size of the source in the present reaction \cite{bay96} 
the classical Gamov correction factor is not appropriate \cite{alb95}. 
For pions the Coulomb correction can be approximately
determined from the experimental correlation function for unlike-sign 
pairs \cite{app98}, which is dominated by the Coulomb interaction and 
receives no contribution from BE correlation. In contrast, the strong 
interaction  cannot be neglected \cite{led82} for unlike-sign kaons. 
Therefore, a parameterization of the Coulomb repulsion factor derived 
from a finite-size source Coulomb wave calculation \cite{sin98} is used 
for the present analysis:
\begin{eqnarray} \label{coul}
A(q) = \tilde{A_c}(r^{\ast},q) ,
\end{eqnarray}
where $r^{\ast}$ is the average distance between the emission points 
of the kaons in the pair center-of-mass system. Following \cite{sin98} 
the value of $r^{\ast}$ can be 
calculated from the correlation radii $R_i$ in Eqs.~\ref{ykp} and \ref{bp}. 
As a consistency check the analogous procedure is applied below to the 
$K^+K^-$ correlations which are also measured in this experiment.

\begin{figure} 
 \begin{center} 
  \leavevmode 
  \epsfxsize=14cm 
  \epsfbox{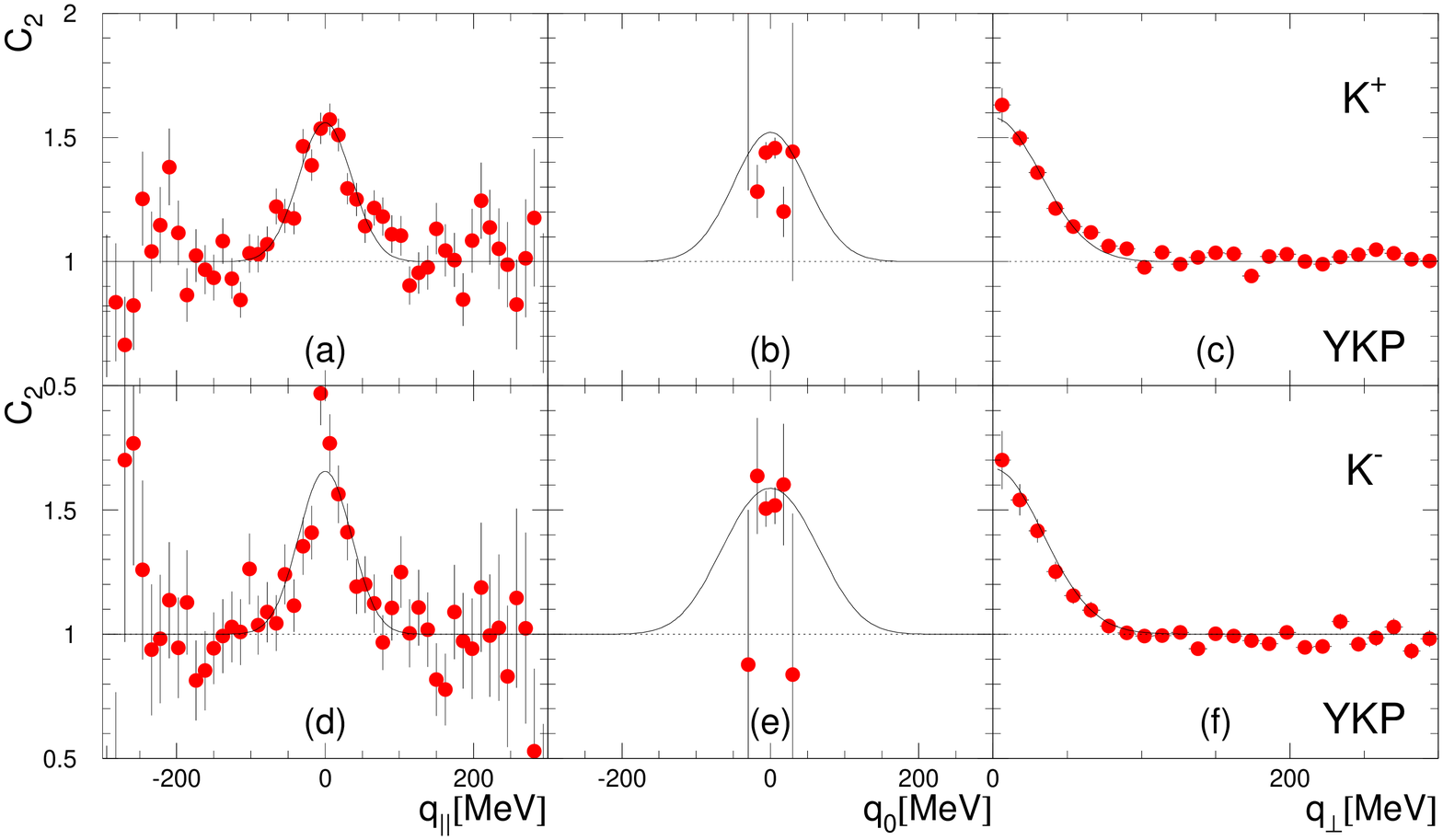} 
  \epsfxsize=14cm 
  \epsfbox{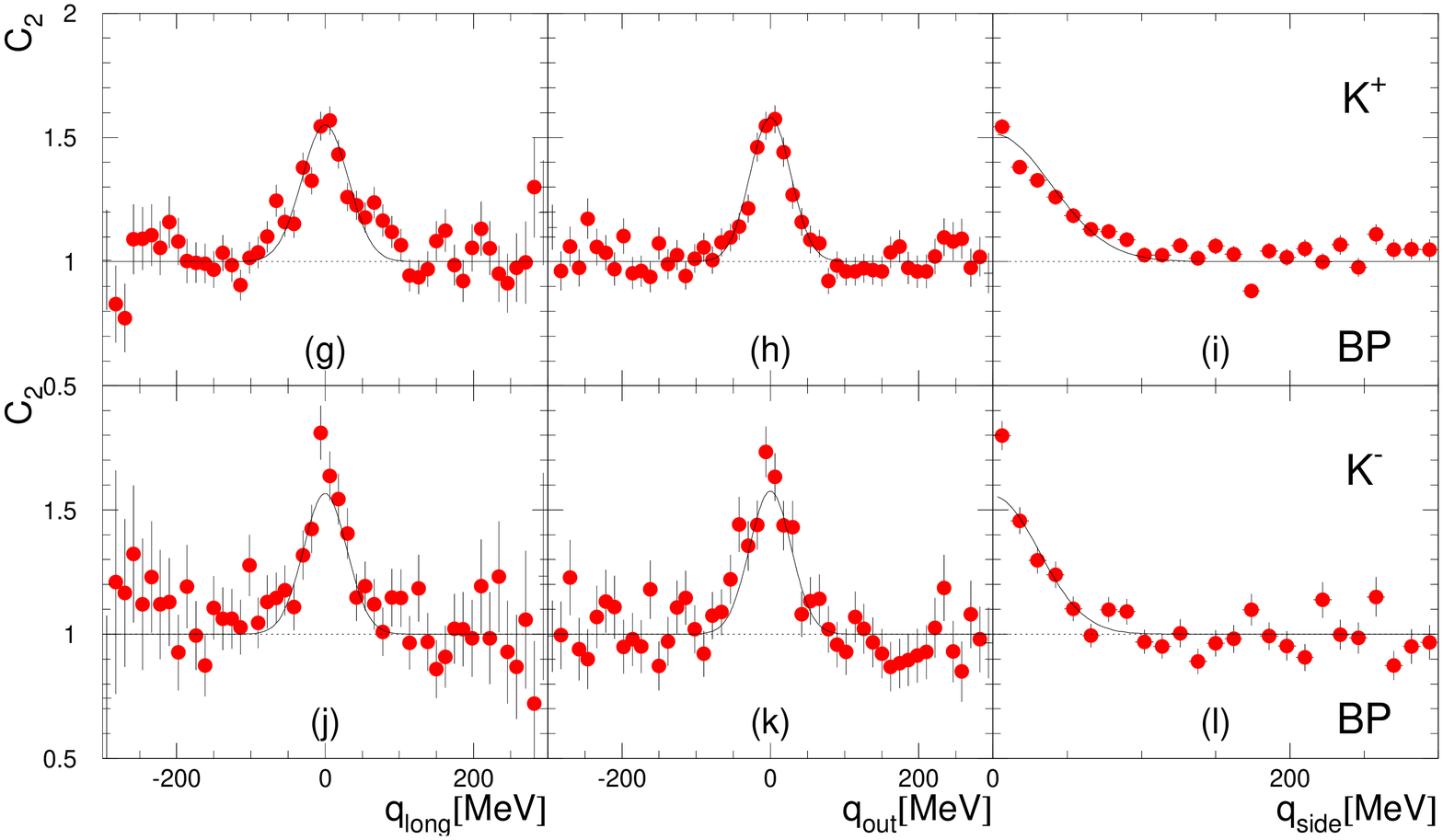} 
 \end{center} 
 \caption{1-dimensional projections on $q_i$ of the Coulomb- and 
purity-corrected 3-dimensional correlation functions Eq.~\ref{corr} 
of like-sign kaons with $q_j\le36$ MeV ($j\ne i$) in YKP (top) and 
BP (bottom) parameterizations. The solid lines in each histogram 
indicate the projections of the fitted 3-dimensional functions of 
Eqs.~\ref{ykp} and ~\ref{bp} with the parameters given in 
Table~\ref{tab_result}.} 
 \label{figure4} 
 \vspace{0.5cm} 
\end{figure} 

The determination of the radius parameters $R_i$ proceeds in an 
iterative way. Starting with estimated values for $R_i$, the Coulomb 
repulsion factor $A(q)$ (Eq.~\ref{coul}) is calculated and smeared with 
the measuring resolution of $q$. A set of $R_i$ and $\lambda$ are then 
obtained from a 3-dimensional maximum likelihood fit of the r.h.s. of 
Eq.~\ref{fitfunc} to the raw correlation function Eq.~\ref{c2_define}. 
The procedure is iterated until the fit parameters reach stable values.

For the central collisions the final results for the YKP and BP 
parameterizations are summarized in Table \ref{tab_result} with the 
fit quality given by $\chi^2/NDF$. The quoted errors are only 
statistical. The systematic errors on the the fitted parameters due 
to uncertainty in Coulomb correction and strong interaction are 
estimated to be about 10\% of each parameter.

\begin{figure} 
 \begin{center} 
  \leavevmode 
  \epsfxsize=9cm 
  \epsfbox{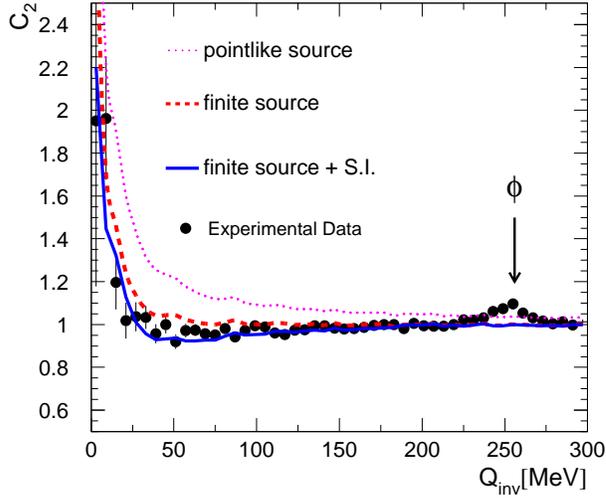} 
 \end{center} 
 \caption{$K^+K^-$ correlation function in central collision events 
plotted versus $Q_{inv}$. The curves show theoretical calculations
(see text) with the effects of measuring resolution and misidentification
applied. The dotted curve represents a pointlike source (Gamov factor). 
The finite source size Coulomb wave calculation is indicated by the 
dashed curve. The final result including also the effect of strong 
interactions is given by the full curve.} 
 \label{figure5} 
 \vspace{0.5cm} 
\end{figure} 

\begin{table*}
  \begin{center} 
  \caption{\label{tab_result} Source parameters with statistical errors 
obtained from 3-dim fits of the central data ; the 1-dim projections 
are shown in Fig.~\ref{figure4}.}   
    \begin{tabular}{c|c|c|c|c|c||c} 
      \hline 
      \hline 
      YKP & $\lambda$ & $R_{\parallel}[fm]$ & $R_{\perp}[fm]$ 
& $R_0[fm]$ & $\beta_{YK}$ & $\chi^2/NDF$ \\ 
      \hline 
      K$^+$ & 0.74$\pm$0.06 & 3.95$\pm$0.35 & 4.34$\pm$0.30 
& 2.86$\pm$0.65 & 0.01$\pm$0.02 & 216/228\\ 
      K$^-$ & 0.86$\pm$0.05 & 4.52$\pm$0.45 & 4.15$\pm$0.28 
& 3.00$\pm$0.52 & 0.01$\pm$0.05 & 270/328\\ 
      \hline 
      \hline 
      BP & $\lambda$ & $R_{long}[fm]$ & $R_{side}[fm]$ 
& $R_{out}[fm]$ & & $\chi^2/NDF$ \\ 
      \hline 
      K$^+$ & 0.81$\pm$0.06 & 4.46$\pm$0.25 & 3.58$\pm$0.40 
& 5.07$\pm$0.27 & & 439/446 \\ 
      K$^-$ & 0.89$\pm$0.07 & 4.78$\pm$0.33 & 4.55$\pm$0.31 
& 4.97$\pm$0.39 & & 499/537 \\ 
      \hline 
      \hline 
    \end{tabular} 
  \end{center} 
  \vspace{0.3cm} 
\end{table*} 

Using Eq.~\ref{fitfunc} and the fitted parameters one can calculate a 
Coulomb corrected correlation function $C_2^{corr}$ from the directly 
measured raw correlation function $C_2^{raw}$ :
\begin{eqnarray} \label{corr}
C_2^{corr}(q) = \frac{ C_2^{raw}(q) - 1  + p}{p\cdot A(q)}           . 
\end{eqnarray} 
The result is shown in Fig.~\ref{figure4} as data points and compared 
to the fitted correlation functions which are depicted by the curves. 
One observes good agreement.

In addition the kaons from about 2 million semi-central (5\% - 20\%) 
events were also analysed and the extracted parameters are listed in 
Table ~\ref{tab_result2}. Both the values of the radius parameters 
and of the correlation intensity are found to be smaller than for the 
central collision events, in qualitative agreement with the expected 
decrease of the radii with the decreasing centrality.

Fig.~\ref{figure5} displays a plot of the $K^+K^-$ correlation 
function versus $Q_{inv}=\sqrt{(\Delta \vec{p})^2-(\Delta E)^2}$ 
for central collision events. The observed correlation is due 
to the Coulomb attraction as well as the strong interaction 
of the $K^+$ and $K^-$ \cite{sin98}. The data points are compared
to theoretical calculations for which the effects of momentum resolution
and misidentification were taken into account. As demonstrated by 
the dotted curve, the Gamov factor (appropriate for a point 
like source) does not describe the data. The finite size Coulomb 
wave calculation using the fitted radius parameters is shown by the dashed 
curve. It already provides a much better representation, but still stays 
somewhat above the data. When also including the contribution from 
strong interactions (full curve) a good reproduction of the region 
$Q_{inv} < 200$ MeV is obtained. This agreement can be taken as a 
confirmation of the Coulomb correction procedure used for like-charge 
kaon correlations.

\begin{table*}  
  \begin{center} 
  \caption{\label{tab_result2} Source parameters from the correlation 
functions of kaons from semi-central events.}   
    \begin{tabular}{c|c|c|c|c|c||c} 
      \hline 
      \hline 
      YKP & $\lambda$ & $R_{\parallel}[fm]$ & $R_{\perp}[fm]$ 
& $R_0[fm]$ & $\beta_{YK}$ & $\chi^2/NDF$ \\ 
      \hline 
      K$^+$ & 0.55$\pm$0.05 & 3.36$\pm$0.38 & 2.96$\pm$0.18 
& 3.41$\pm$0.72 & 0.02$\pm$0.02 & 2573/2019\\ 
      K$^-$ & 0.60$\pm$0.06 & 4.07$\pm$0.51 & 3.95$\pm$0.38 
& 2.03$\pm$0.82 & 0.01$\pm$0.04 & 1553/1574\\ 
      \hline 
      \hline 
      BP & $\lambda$ & $R_{long}[fm]$ & $R_{side}[fm]$ 
& $R_{out}[fm]$ & & $\chi^2/NDF$ \\ 
      \hline 
      K$^+$ & 0.60$\pm$0.03 & 3.35$\pm$0.35 & 3.16$\pm$0.48 
& 3.40$\pm$0.26 & & 1529/2012 \\ 
      K$^-$ & 0.67$\pm$0.04 & 3.55$\pm$0.50 & 3.36$\pm$0.51 
& 3.95$\pm$0.37 & & 968/769 \\ 
      \hline 
      \hline 
    \end{tabular} 
  \end{center} 
 \vspace{0.3cm} 
\end{table*}

\section{Discussion and Conclusion} \label{sec:dis} 

From Tables \ref{tab_result} and \ref{tab_result2} 
one observes that the radius parameters for K$^+$ and K$^-$ are
similar with a trend to somewhat larger values for K$^-$.
This is consistent with the 
expected influence of Coulomb interaction with the positive charge of 
the reaction zone \cite{bay96}. The correlation intensity parameter 
$\lambda$ stays below unity. Possible causes are e.g. a partially 
coherent source, a contribution to the observed kaons from decays 
of long-lived resonances, or a deviation from the Gaussian 
parameterization due to a mixture of sources with different 
radii \cite{led79}. The longitudinal source velocities $\beta_{YK}$ 
are close to zero in agreement with the fact that our acceptance is 
restricted to midrapidity (see Fig.~\ref{figure2}b).

\begin{figure} 
 \begin{center} 
  \leavevmode 
  \epsfxsize=14cm 
  \epsfbox{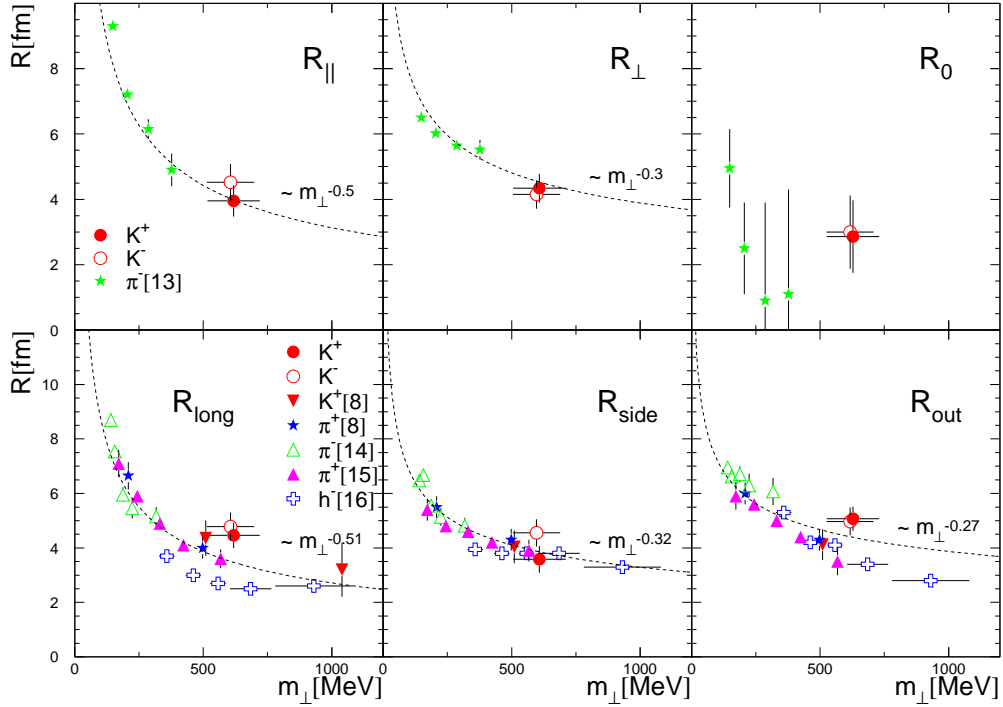} 
 \end{center} 
 \caption{Comparison of radius parameters of kaons and pions as 
function of $m_\perp$. Error bars show statistical and systematic errors
added in quadrature. The dashed lines indicate the fit of the function 
$Cm_\perp^{-\alpha}$ with different $\alpha$, respectively.} 
 \label{figure6} 
\end{figure} 

 When comparing the radius parameters obtained from kaon 
correlations with those of pions one has to bear in mind that these 
parameters depend on the transverse momentum or transverse 
mass as a consequence of the collective expansion of the source 
\cite{hei96b}. In Fig. \ref{figure6} we show the earlier pion data 
\cite{bea01,app98,ros02,ada02,ant01} for the same reaction  together with 
the kaon results from Table \ref{tab_result} as well as those from the 
CERN experiment NA44 \cite{bea01}. The radii show $m_\perp^{-\alpha}$ 
dependence with  $\alpha\approx 0.5$ for the longitudinal radii 
(as suggested by Makhlin and Sinyukov \cite{mak88}) and 
$\alpha\simeq 0.3$ for the transverse radii. 
 
\begin{figure} 
 \begin{center} 
  \leavevmode 
  \epsfxsize=14cm 
  \epsfbox{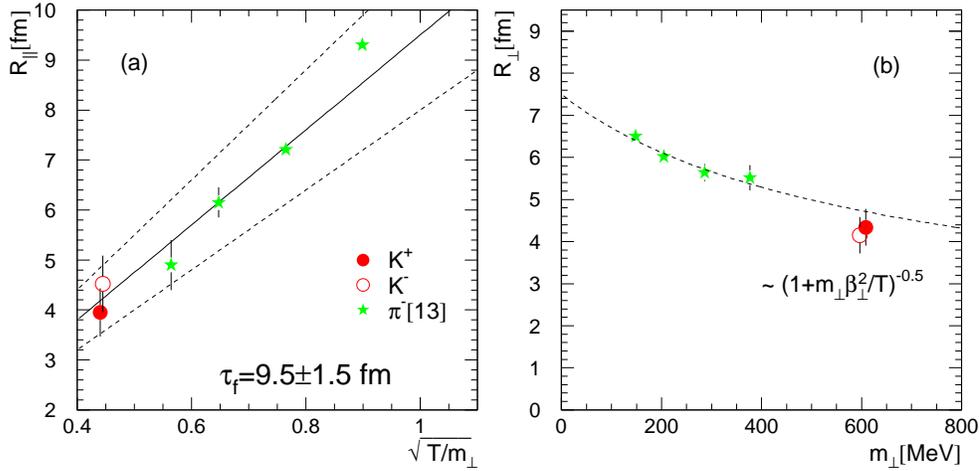} 
 \end{center} 
 \caption{(a) $R_\parallel$ as a function of $\sqrt{T/m_\perp}$ 
according to Eq.~\ref{tau}. The solid line indicates the function 
with $\tau_f=9.5$ fm with $\pm$1.5 fm indicated by the dashed lines. 
(b) $R_\perp$ as function of $m_\perp$. The dashed line denotes a 
model calculation \cite{cha95,app98} with the transverse expansion 
velocity $\beta_\perp=0.55$.} 
 \label{figure7} 
 \vspace{0.5cm} 
\end{figure}

  According to the relation : 
\begin{eqnarray} \label{tau} 
R_\parallel=\tau_f\sqrt{\frac{T}{m_\perp}}, 
\end{eqnarray} 
based on a hydrodynamic expansion model \cite{mak88}, where $\tau_f$ 
and $T$ are the freeze-out time and temperature of the particles, 
respectively, the longitudinal radii can be plotted as a function of 
$\sqrt{T/m_\perp}$ (Fig.~\ref{figure7}a) using the known freeze-out 
temperature $T\simeq 120$ MeV from \cite{app98}.  
 From its gradient the freeze-out time is extracted as $9.5\pm1.5$ fm, 
which is consistent with the previous estimate based on the pion correlation 
study in \cite{app98}. Furthermore, the relatively weak dependence of 
$R_\perp$ on $m_\perp$ can also be reproduced by hydrodynamic 
expansion models \cite{cha95,app98} with a transverse expansion 
velocity $\beta_\perp$ around 0.55 (see Fig.~\ref{figure7}b).

  In conclusion the kaon radius parameters are fully consistent with 
the published pion results and the hydrodynamic expansion model. 
The influence of different contributions of resonance formation and 
different rescattering cross sections seem to be either minor or similar 
in both cases. Pions and kaons thus seem to decouple simultaneously.

\section*{Acknowledgements} 
Acknowledgements: This work was supported by the Director, 
Office of Energy Research, Division of Nuclear Physics of the 
Office of High Energy and Nuclear Physics of the US Department 
of Energy (DE-ACO3-76SFOOO98 and DE-FG02-91ER40609), 
the US National Science Foundation, the Bundesministerium fur 
Bildung und Forschung, Germany, the Alexander von Humboldt 
Foundation, the UK Engineering and Physical Sciences Research 
Council, the Polish State Committee for Scientific Research 
(2 P03B 130 23 and 2 P03B 02418), the Hungarian Scientific 
Research Foundation (T14920, T32293 and T032648), Hungarian National 
Science Foundation, OTKA, (F034707), the EC Marie Curie Foundation, 
and the Polish-German Foundation.

\end{document}